\documentstyle[art12,epsf]{article}
\catcode`\@=11

\catcode`\@12
\begin{document}
\pagestyle{myheadings}
\typeout{}
\newcommand{\eqn}[1]{(\ref{eq:#1})}
\newcommand{\ben}{\begin{equation}}
\newcommand{\een}{\end{equation}}
\newcommand{\bea}{\begin{eqnarray}}
\newcommand{\eea}{\end{eqnarray}}
\newcommand{\nn}{\nonumber \\ }
\newcommand{\bdm}{\begin{displaymath}}
\newcommand{\edm}{\end{displaymath}}
\newcommand{\pa}{\partial }
\newcommand{\hf}{\frac{1}{2}}
\newcommand{\slr}{sl(2/1;{\bf R})}
\newcommand{\slc}{sl(2/1;{\bf C})}
\newcommand{\hslr}{\hat{sl}(2/1;{\bf R})}
\newcommand{\hslc}{\hat{sl}(2/1;{\bf C})}
\newcommand{\Slr}{SL(2/1;{\bf R})}
\newcommand{\rt}{\sqrt{2k}}
\newcommand{\irt}{\frac{1}{\sqrt{2k}}}
\newcommand{\frt}{\sqrt{\frac{k}{2}}}
\newcommand{\ifrt}{\sqrt{\frac{2}{k}}}
\newcommand{\iap}{i\alpha_+}
\newcommand{\pap}{\psi'_1}
\newcommand{\pbp}{\psi'_2}
\newcommand{\psa}{\psi_1}
\newcommand{\psb}{\psi_2}
\newcommand{\papd}{\psi^{'\dagger}_1}
\newcommand{\pbpd}{\psi^{'\dagger}_2}
\newcommand{\fm}{\phi_-}
\newcommand{\fp}{\phi_+}
\newcommand{\tpap}{\tilde{\psi}'_1}
\newcommand{\tpbp}{\tilde{\psi}'_2}
\newcommand{\tpsa}{\tilde{\psi}_1}
\newcommand{\tpsb}{\tilde{\psi}_2}
\newcommand{\tpapd}{\tilde{\psi}^{'\dagger}_1}
\newcommand{\tpbpd}{\tilde{\psi}^{'\dagger}_2}
\newcommand{\tfm}{\tilde{\phi}_-}
\newcommand{\tfp}{\tilde{\phi}_+}
\newcommand{\Bp}{\exp[\tilde{\phi}\cdot{\bf h}]}
\newcommand{\Bpi}{\exp[-\tilde{\phi}\cdot{\bf h}]}
\parskip 2ex

\begin{titlepage}
\begin{flushright}
DTP/96/21\\
May 1996\\
\end{flushright}
\vspace{1cm}
\begin{center}
{\Large\bf  Representation theory of the affine 
Lie superalgebra $\hslc$ at fractional level.}\\
\vspace{1cm}
{\large
P. Bowcock and A. Taormina}\\
\vspace{0.5cm}
{\it Department of Mathematical Sciences, University of Durham,\\
Durham DH1~3LE, England} \\
\vspace{0.5cm}
\end{center}
\begin{abstract} 
{}N=2 noncritical strings are closely related to the $\Slr/\Slr$ 
Wess-Zumino-
Novikov-Witten model, and there is much hope to further probe the former by 
using the algebraic apparatus provided by the latter. An important ingredient
is the precise knowledge of the $\hslc$ representation theory at fractional 
level. In this paper, 
the embedding diagrams of singular vectors appearing in $\hslc$
Verma modules for fractional values of the level ($k=p/q-1$, p and q coprime)
are derived analytically. The nilpotency of the fermionic generators in
$\hslc$ requires the introduction of a nontrivial 
generalisation of the MFF construction to relate singular vectors among 
themselves. The diagrams reveal a striking similarity with the
degenerate representations of the $N=2$ superconformal algebra.
\end{abstract}
\vskip 6truecm
{}e-mail addresses : Peter.Bowcock@durham.ac.uk , Anne.Taormina@durham.ac.uk
\end{titlepage}

\section{Introduction}

{}The $N=2$ 
noncritical string possesses interesting features and technical 
challenges. In particular, as emphasized in \cite{distler},
this string theory is not confined to the regime of weak 
gravity, i.e. the phase transition point between weak and strong gravity 
regimes is not of the same nature as in the $N=0,1$ cases. This absence of 
barrier in the central charge is source of complications, but also the hope 
of some new physics. 
The focus on noncritical strings in recent years was initially motivated by
the nonperturbative definition of string theory
in space-time dimension $d < 1$ in the context of matrix models.
The continuum approach, which involves the quantisation 
of the Liouville theory, gives results which are in agreement with those
obtained in matrix models, on the scaling behaviour of correlators for instance
\cite{goulian}. Although less powerful, the continuum approach generalises to 
supersymmetric strings. Some effort has been put in the study of $N=1$ and $N=2$
noncritical superstrings, but no clear picture has emerged so far as how useful
they might be, in particular in extracting nonperturbative information
\cite{distler,fy2,abdalla1,anton,abdalla2}.

A particularly promising tool in the description and understanding of 
noncritical (super)strings
is the use of gauged
$G/G$ Wess-Zumino-Novikov-Witten (WZNW) models, with $G$ a Lie (super)group,
\cite{yankiel,huyu,fy1,fy2}. For instance,
the $\Slr /\Slr $ topological quantum field theory obtained by gauging 
the
anomaly free diagonal subgroup $\Slr$ of the global $\Slr _L \times \Slr _R$
symmetry of the WZNW model appears to be 
intimately related to the noncritical charged fermionic string, which 
is the prototype of $N=2$ supergravity in two dimensions.  
A comparison of the ghost content of the two theories strongly 
suggests that the $N=2$ noncritical string is equivalent to the 
tensor product of a $\em twisted$ $\Slr /\Slr $ WZNW model with 
the topological theory of a spin $1/2$ system \cite{us}. 
It is however only when a one-to-one correspondence 
between the physical states and equivalence of the correlation functions 
of the two theories are established that one can view the twisted $G/G$ 
model as the topological version of the corresponding noncritical string 
theory. For the bosonic string, 
the recent derivation of conformal blocks for admissible representations
of $\widehat{sl(2;{\bf R})}$ is a major step in this direction \cite{jens}.

Our aim in this paper is to provide the algebraic background for the analysis 
of the $N=2$ noncritical string viewed as the $\Slr /\Slr $ WZNW theory.
The physical states of the latter theory will be obtained in a 
forthcoming publication as elements of the cohomology of the BRST 
charge \cite{us}. The procedure we follow is by now quite standard 
\cite{bouw,yankiel,fy1,fy2}. The partition function of the 
$\Slr /\Slr $ theory splits in three sectors : a level $k$  
and a level $-(k+2)$ WZNW models based on $\Slr$
as well as  a system of four fermionic ghosts 
$(b_a,c^a), a=\pm , 3,4$ and four bosonic ghosts 
$(\beta_{\alpha},\gamma^{\alpha}),
(\beta'_{\alpha}, \gamma'_{\alpha}),\alpha =\pm \hf$
corresponding to the four even (resp. odd) generators of $\Slr$ 
\cite{PW,ks,gk,yankiel}.

The cohomology is calculated on the space 
${\cal{F}} _k \otimes {\cal{F}}_{-(k+2)}
\otimes {\cal {F}}_2$ where ${\cal{F}}_k$ denotes the space of 
irreducible representations of $\hslr_k$, while 
${\cal{F}}_{-(k+2)}$ and ${\cal{F}}_2$ denote the Fock spaces of 
the level $-(k+2)$ and ghosts sectors respectively. As a first step, 
one calculates the cohomology on the whole Fock space, using a 
free field  representation of $\hslr$ and its dual. These are the 
Wakimoto modules constructed in \cite{us2}. 
In a second step, one must pass from the cohomology on the Fock space 
to the irreducible representations of $\hslc$ at fractional level $k$.  
After we review some basic properties of $\slr$ in Section 2,
we give in Section 3 the $\hslc$ generalisation of the Malikov-Feigin-Fuchs
construction needed to relate the bosonic and fermionic singular vectors
whose quantum numbers are derived from the generalised Kac-Kazhdan
determinant formula \cite{k1}. We stress here that the superalgebra $\slr$ has 
nilpotent fermionic generators, or, in other words, has lightlike fermionic 
roots.
This property is the source of several interesting algebraic complications
in the study of the representation theory of its affine counterpart, in 
contrast for instance with the well-studied -and much simpler case- of 
$\widehat{Osp(1/2)}$, relevant for $N=1$ noncritical strings
\cite{kw,fy1,fy2}.
Section 4 provides a classification of embedding diagrams for singular vectors
appearing in highest weight Verma modules of $\hslc$ at fractional level
$k=p/q-1$, for $p$ and $q$ coprime. The four classes are determined by
the highest weight states quantum numbers, given by the zeros of the 
generalised Kac-Kazhdan determinant. These diagrams are characterized by the 
fact that they contain an 
infinite number of singular vectors, and, according to the general theory of 
Kac and Wakimoto \cite{kw}, admissible representations should be a subset
of the irreducible representations obtained as cosets of the Verma modules by 
the singular modules. Admissible representations, although generically 
non integrable, have characters which transform as finite representations
of the modular group. They are ultimately the representations needed to 
derive the space of physical states of the $\Slr /\Slr $ WZNW model.

\section{The Lie superalgebra $\slr$: a brief review}

The set $\cal{M}$ of $3 \times 3$ matrices with real entries $m_{ij}$
whose diagonal elements satisfy the super-tracelessness condition
\ben
m_{11}+m_{22}-m_{33}=0
\een
forms, with the standard laws of matrix addition and multiplication,
the real Lie superalgebra $\slr$. Any matrix ${\bf m} \in \cal{M}$ can
be expressed as a real linear combination of eight basis matrices

\bea
{\bf m}&=&m_{11}{\bf h_1}+m_{22}{\bf h_2}+m_{12}{\bf e_{\alpha_1 +\alpha_2}} 
       +m_{21}{\bf e_{-(\alpha_1 +\alpha_2)}}\nn
       &+&m_{32}{\bf e_{\alpha_1}}+m_{23}{\bf e_{-\alpha_1}}
       + m_{13}{\bf e_{\alpha_2}}+m_{31}{\bf e_{-\alpha_2}}
\eea
with
\bea
{\bf h}_1 = \left(
\begin{array}{cccc}
1 & 0 & \vrule & 0 \\
0 & 0 & \vrule & 0 \\
\hline 
0 &  0 & \vrule & 1 \\
\end{array}
\right ),&&
{\bf h}_2 = \left(
\begin{array}{cccc}
0 & 0 & \vrule & 0 \\
0 & 1 & \vrule & 0 \\
\hline 
0 & 0 & \vrule & 1 \\
\end{array}
\right ),\nonumber \\
{\bf e}_{\alpha_1+\alpha_2} = \left(
\begin{array}{cccc}
0 & 1 & \vrule & 0 \\
0 & 0 & \vrule & 0 \\
\hline 
0 & 0 & \vrule & 0 \\
\end{array}
\right ),&&
{\bf e}_{-(\alpha_1+\alpha_2)}=\left(
\begin{array}{cccc}
0 & 0 & \vrule & 0 \\
1 & 0 & \vrule & 0 \\
\hline 
0 & 0 & \vrule & 0 \\
\end{array}
\right ),\nonumber \\
{\bf e}_{\alpha_1}=\left(
\begin{array}{cccc}
0 & 0 & \vrule & 0\\
0 & 0 & \vrule & 0 \\
\hline 
0 & 1 & \vrule & 0 \\
\end{array}
\right ),&&
{\bf e}_{-\alpha_1}=\left(
\begin{array}{cccc}
0 & 0 & \vrule & 0 \\
0 & 0 & \vrule & 1 \\
\hline 
0 & 0 & \vrule & 0 \\
\end{array}
\right ),\nonumber \\
{\bf e}_{\alpha_2}=\left(
\begin{array}{cccc}
0 & 0 & \vrule & 1 \\
0 & 0 & \vrule & 0 \\
\hline 
0 & 0 & \vrule & 0 \\
\end{array}
\right ),&&
{\bf e}_{-\alpha_2}=\left(
\begin{array}{cccc}
0 & 0 & \vrule & 0 \\
0 & 0 & \vrule & 0 \\
\hline 
1 & 0 & \vrule & 0 \\
\end{array}
\right ).
\label{eq:fr}\eea
One can associate a $Z_2$ grading to these basis matrices by
partitioning them into four submatrices of dimensions $2 \times 2, 2
\times 1, 1\times 2 $ and $1 \times 1$, and calling even (resp. odd)
those with zero off-diagonal (resp. diagonal) submatrices. From this
fundamental 3-dimensional representation of $\slr$, one can write down
the (anti)-commutation relations obeyed by its four bosonic generators
${H_{\pm},E_{\pm(\alpha_1+\alpha_2)}}$ (corresponding to the even
basis matrices ${\bf h_{\pm}}={\bf h_1} \pm {\bf h_2}, {\bf
e_{\pm(\alpha_1 +\alpha_2)}}$) and its four fermionic generators
$E_{\pm \alpha_1}, E_{\pm \alpha_2}$ (corresponding to the odd basis
matrices),
\bea
&&[E_{\alpha_1+\alpha_2},E_{-(\alpha_1+\alpha_2)}]=H_1-H_2~~~,~~~
[H_1-H_2,E_{\pm(\alpha_1+\alpha_2)}]=\pm 2E_{\pm(\alpha_1+\alpha_2)},\nn
&&[E_{\pm(\alpha_1+\alpha_2)},E_{\mp \alpha_1}]=\pm E_{\pm \alpha_2}~~~,~~~
[E_{\pm(\alpha_1+\alpha_2)},E_{\mp \alpha_2}]=\mp E_{\pm \alpha_1},\nn
&&[H_1-H_2,E_{\pm \alpha_1}]=\pm  E_{\pm \alpha_1}~~~,~~~
[H_1-H_2,E_{\pm \alpha_2}]=\pm  E_{\pm \alpha_2},\nn
&&[H_1+H_2,E_{\pm \alpha_1}]=\pm E_{\pm \alpha_1}~~~,~~~
[H_1+H_2,E_{\pm \alpha_2}]=\mp  E_{\pm \alpha_2},\nn
&&\{ E_{\alpha_1},E_{-\alpha_1}\}=H_2~~~,~~~
\{ E_{\alpha_2},E_{-\alpha_2}\}=H_1~~~,~~~
\{E_{\pm\alpha_1},E_{\pm \alpha_2}\}=E_{\pm(\alpha_1+\alpha_2)}.
\label{eq:cr}\eea

As can be seen from the above commutation relations ,
the even subalgebra of $\slr$ is the direct sum of an abelian algebra 
generated by
$H_+=H_1+H_2$ and of the real Lie algebra $sl(2;{\bf R})$ generated by 
$E_{\pm (\alpha_1+\alpha_2)}$ and $H_-=H_1-H_2$. 
Because the semi-simple
part  $sl(2;{\bf R})$ of its even subalgebra is noncompact,
the Lie superalgebra $\slr$ is a noncompact form of its complexification
$A(1,0)$. We follow here the notations of Kac \cite{kac77}. According to 
Parker \cite{parker}, the three real forms of $A(1,0)$ are $\slr$,
$su(1,1/1)$ and $su(2/1)$. Although the latter is actually the compact form
of $A(1,0)$, the corresponding Lie supergroup is noncompact \cite{cornwell}.
The finite dimensional irreducible representations of $\slr$,
which incidentally is isomorphic to $osp(2/2;{\bf R})$, are constructed in 
\cite{scheu} and in \cite{dhoker}. 
A construction of oscillator-like unitary irreducible 
representations of $\slr$ is given in \cite{bars}.
The Cartan-Killing metric is given by a quadratic expression in the
structure constants. It generalises the purely bosonic case in
incorporating the $Z_2$ grading by associating degree zero to the
bosonic generators, and degree 1 to the fermionic ones,
\bea
g_{\alpha \beta}&=&f_{\alpha \gamma}^{~~~\rho}f_{\beta
\rho}^{~~~\gamma}(-1)^{d(\rho)}\nn
d(\rho )&=&0~~~~~\rm{for~ \rho~a~bosonic ~index}\nn
d(\rho )&=&1~~~~~\rm{for~ \rho~a~fermionic ~index}.
\eea
The bosonic indices take the values $\pm,\pm(\alpha_1+\alpha_2)$, 
while the fermionic indices take the values
$\pm \alpha_1,\pm \alpha_2$. One explicitly has
\bea
g_{--}=-g_{++}=1 &,& g_{\alpha_1+\alpha_2,-(\alpha_1+\alpha_2)}=
g_{-(\alpha_1+\alpha_2),\alpha_1+\alpha_2}=2\nn
g_{\alpha_1,-\alpha_1}=-g_{-\alpha_1,\alpha_1}=-2&,&
g_{\alpha_2,-\alpha_2}=-g_{-\alpha_2,\alpha_2}=2.
\eea
The quadratic Casimir is given by,
\bea
C^{(2)}&=&\frac{1}{2}((H_--H_+)(H_-+H_+)+E_{\alpha_1 +\alpha _2}
E_{-(\alpha_1 +\alpha _2)}+E_{-(\alpha_1 +\alpha _2)}E_{\alpha_1 +\alpha _2}\nn
&&~~~~+E_{\alpha_1}E_{-\alpha_1}-E_{-\alpha_1}E_{\alpha_1}
-E_{\alpha_2}E_{-\alpha_2}+E_{-\alpha_2}E_{\alpha_2}),
\label{eq:C2}\eea
and the atypical representations are those for which the quadratic Casimir 
vanishes.
The fermionic nonzero roots $\pm \alpha_1,\pm \alpha_2$ have length 
square zero, and we normalise the bosonic nonzero roots 
$\pm (\alpha_1 +\alpha_2)$ 
to have length square 2. The root diagram can be represented
in a 2-dimensional Minkowski space with the fermionic roots in the
lightlike directions
\vskip .5cm

\setlength{\unitlength}{0.012500in}%
\begingroup\makeatletter\ifx\SetFigFont\undefined
\def\x#1#2#3#4#5#6#7\relax{\def\x{#1#2#3#4#5#6}}%
\expandafter\x\fmtname xxxxxx\relax \def\y{splain}%
\ifx\x\y   
\gdef\SetFigFont#1#2#3{%
  \ifnum #1<17\tiny\else \ifnum #1<20\small\else
  \ifnum #1<24\normalsize\else \ifnum #1<29\large\else
  \ifnum #1<34\Large\else \ifnum #1<41\LARGE\else
     \huge\fi\fi\fi\fi\fi\fi
  \csname #3\endcsname}%
\else
\gdef\SetFigFont#1#2#3{\begingroup
  \count@#1\relax \ifnum 25<\count@\count@25\fi
  \def\x{\endgroup\@setsize\SetFigFont{#2pt}}%
  \expandafter\x
    \csname \romannumeral\the\count@ pt\expandafter\endcsname
    \csname @\romannumeral\the\count@ pt\endcsname
  \csname #3\endcsname}%
\fi
\fi\endgroup
\begin{picture}(151,180)(017,435)
\thicklines
\multiput(180,603)(16.64190,-22.18921){8}{\line( 3,-4){  8.667}}
\put(305,436){\vector( 3,-4){0}}
\put(180,603){\vector(-3, 4){0}}
\multiput(305,603)(-16.64190,-22.18921){8}{\line(-3,-4){  8.667}}
\put(180,436){\vector(-3,-4){0}}
\put(305,603){\vector( 3, 4){0}}
\put(117,520){\vector(-1, 0){  0}}
\put(117,520){\vector( 1, 0){251}}
\put(243,520){\circle*{6}}
\put(315,435){\makebox(0,0)[lb]{\smash{${\bf \alpha_2}$}}}
\put(315,600){\makebox(0,0)[lb]{\smash{${\bf \alpha_1}$}}}
\end{picture}
\vskip 1cm
\centerline{{\bf {\rm Fig.1}}:\it ~The root diagram of $A(1,0)$}

\noindent

The Weyl group of $\slr$ is isomorphic to the Weyl group of its even
simple subalgebra $sl(2;{\bf R})$. There is no obvious concept of a
Weyl reflection about the hyperplane orthogonal to a zero square norm
fermionic root.  If one therefore chooses a purely fermionic system of
simple roots $\{\alpha_1,\alpha_2 \}$, there is no element of the Weyl
group which can transform it into the system of simple roots
$\{-\alpha_2,\alpha_1+\alpha_2\}$. Dobrev and Petkova \cite{dp} and 
later, Penkov and Serganova \cite{serga} have actually extended
the definition of the Weyl group to incorporate the transformation
$\alpha_2 \rightarrow -\alpha_2$.  This non uniqueness of
the generalized Dynkin diagram for Lie superalgebras is well
established \cite{kw2}.

\section{Generalisation of the Malikov-Feigin-Fuchs construction}

As pointed out in the introduction, the physical spectrum of the 
$\Slr /\Slr$ WZNW topological model is determined by the structure 
of the Wakimoto modules given in \cite{us2}, and by the 
structure of irreducible representations of the affine superalgebra 
$A(1,0)^{(1)} \equiv \hslc$
at fractional level $k$.

As first discussed in \cite{kw,kw2}, integrable highest weight 
state representations of $A(1,0)^{(1)}$ require the level $k$ to 
be integer. The corresponding Verma modules, when reducible, 
contain an infinite number of singular vectors, and the characters 
of the associated irreducible representations are asserted to form a finite 
dimensional representation of the modular group in \cite{kw}.
If one relaxes the condition $k \in {\bf Z}_+$ and allows the level to be 
fractional, the reducible Verma modules still contain an infinite 
number of singular vectors and, for appropriate choices of highest 
weight state quantum numbers, the corresponding irreducible 
representations still have characters written in terms of Theta
functions and are believed to transform 
as finite dimensional representations of the 
modular group. Such irreducible representations are called $\em admissible $, 
according to the terminology introduced by Kac and Wakimoto \cite {kw}, 
and they are precisely the irreducible representations which enter in 
the analysis of the BRST cohomology of the $\Slr /\Slr$ WZNW model.  
In order to understand their structure, we use  
the determinant formula for the contravariant bilinear form associated 
to infinite dimensional contragredient Lie superalgebras. This formula 
is a straight generalisation of the Kac-Kazhdan formula giving the 
determinant of the bilinear form associated to affine algebras \cite{kk}, and 
appears in \cite{k1} and \cite{dob}. A more recent 
derivation is due to \cite{fy1}. It reads,

\ben
{\rm det} F_{\eta}(\Lambda)=
\prod _{n \in {\bf Z}_+ \setminus \{0\} } \prod _{\alpha \in 
\tilde{\Delta}^+_0} 
[\tilde{\phi}_n^{(0)}(\alpha )]^{P(\eta-n\alpha)}
\prod _{n \in {1+ 2{\bf Z}}_+} \prod _{\alpha \in \Delta^+_1€\setminus 
\tilde{\Delta}^+_1} [\phi _n^{(1)}(\alpha )]^{P(\eta-n\alpha)}
 \prod _{\alpha \in \tilde{\Delta}^+_1} 
[\tilde{\phi}^{(1)}(\alpha )]^{P_{\alpha}(\eta-\alpha)},\nn
\een
with
\bea
\tilde{\phi }_n^{(0)}(\alpha )&=&\phi _n^{(1)}(\alpha )=
(\Lambda + \rho , \alpha )-\hf n (\alpha ,\alpha)\nn
{\tilde{\phi }}^{(1)}(\alpha )&=& (\Lambda + \rho , \alpha ).
\eea
We denote by $\Delta $ the full set of roots, $\Delta_{\epsilon}^+$ 
$(\Delta_{\epsilon}^-)$ is the set of positive (negative) even 
$(\epsilon =0)$ and odd $(\epsilon =1)$ roots.
Also, 
\bea
\tilde{\Delta }^+_0&=& \{ \alpha \in \Delta^+_0 : \hf \alpha \notin \Delta\}\nn
{\tilde{\Delta }}^+_1&=& \{ \alpha \in \Delta^+_1 : (\alpha ,\alpha)=0\}. 
\eea

Determinant formulas of the kind above provide powerful information on 
the reducibility of Verma modules with highest weight $\Lambda$. At 
fixed $\Lambda$, there exists an infinity of such formulas, each 
corresponding to an element $\eta $ of $\Gamma _+$, the semi-group 
generated by the positive roots,
\ben
\eta = \sum _{\alpha _i \in \Delta ^+}n_i \alpha _i,
\label{eq:eta}\een
where $n_i \in \{0,1\}$ if $2\alpha_i \in \Delta_0^+$, and 
$n_i \in {\bf Z}_+$ otherwise. The vector $\rho $ is defined as 
\ben 
\rho = (\bar {\rho}, h^{\nu }, 0)
\een
where $h^{\nu}$ is the dual Coxeter number of the corresponding 
finite dimensional Lie superalgebra and $\bar {\rho}$ is half the 
graded sum of its positive roots,
\ben
\bar{\rho}=\hf (\sum_{\bar{\alpha} \in \bar{\Delta}_0^+} \bar{\alpha}-
\sum_{\bar{\alpha} \in \bar{\Delta}_1^+} \bar{\alpha}).
\een
$P(\eta)$ is the number of partitions of $\eta $, i.e. the number of 
ways $\eta$ can be written as a linear combination of positive roots 
with the restrictions described in \eqn{eta}. One sets $P(0)=1$ and $P(\eta )=0$
if $\eta \notin \Gamma _+$. Furthermore, $P_{\alpha }(\eta )$ denotes 
the number of partitions of $\eta $ which do not contain $\alpha $. 
A criterion for irreducibility of Verma modules with highest weight 
$\Lambda $, $V(\Lambda )$, is that  
${\rm det} F_{\eta }(\Lambda )\neq 0~~ \forall \eta \in \Gamma _+$. 
If ${\rm det }F_{\eta }(\Lambda )=0$, the Verma module contains 
singular vectors which generate irreducible submodules. 

Let us now specialise to $A(1,0)^{(1)}$, and extract from the Kac determinant 
the quantum numbers of representations of fractional level. We introduce the 
following Laurent expansions for the $A(1,0)^{(1)}$ currents,
\bea
J({\bf e}_{\pm(\alpha_1+\alpha_2)})(z)&=&
\sum _n J_n^{\pm} z^{-n-1},\nn
J({\bf h}_-)(z)&=&
2\sum _n J_n^3 z^{-n-1},~~~~~~~~~~~ 
J({\bf h}_+)(z)
= 2\sum_n U_n z^{-n-1},\nn
J({\bf e_{\pm \alpha_1}})(z) &=& 
\sum _n j_n^{'\pm}z^{-n-1},~~~~~~~~~~J({\bf e_{\pm \alpha_2}})(z) = 
\sum _n j_n^{\pm}z^{-n-1}. 
\eea
In terms of these Laurent modes, 
the commutation relations for $A(1,0)^{(1)}$ are,
\bea
[J_m^+,J_n^-]&=& 2J_{m+n}^3+km \delta_{m+n,0}\nn
{[} J_m^3,J_n^{\pm} {]} &=& \pm J_{m+n}^{\pm},~~~~~
[J_m^3,J_n^3]=\frac{k}{2}m \delta_{m+n,0}\nn
{[}J_m^{\pm},j_n^{'\mp}{]}&=&\pm j_{m+n}^{\pm},~~~~~
{[}J_m^{\pm},j_n^{\mp}{]}=\mp j_{m+n}^{\pm '}\nn
{[}2J_m^3,j_n^{'\pm}{]}&=&\pm j_{m+n}^{'\pm},~~~~~
[2J_m^3,j_n^{\pm}]=\pm j_{m+n}^{\pm}\nn
{[}2U_m,j_n^{'\pm}{]}&=&\pm j_{m+n}^{'\pm},~~~~~
[2U_m,j_n^{\pm}]=\mp j_{m+n}^{\pm}\nn
{[}U_m,U_n{]}&=& -\frac{k}{2}m \delta_{m+n,0}\nn
\{j_m^{'+},j_n^{'-} \}&=& (U_{m+n}-J_{m+n}^3)-km \delta_{m+n,0}\nn
\{j_m^+,j_n^- \}&=& (U_{m+n}+J_{m+n}^3)+km \delta_{m+n,0}\nn
\{ j_m^{'\pm},j_n^{\pm} \} &=& J_{m+n}^{\pm}.
\label{eq:super}\eea
The Sugawara energy-momentum tensor is given by,
\bea
T(z)&=& \frac{1}{2(k+1)} \{ 2(J^3)^2(z) - 2U^2(z) + J^+J^-(z) + J^-J^+(z)\nn
&& +j^{'+}j^{'-}(z) - j^{'-}j^{'+}(z) 
-j^+j^-(z) + j^-j^+(z) \}.
\eea
Its zero-mode subalgebra possesses an automorphism $\tau $ of order 2,
\bea
&&\tau (J^{\pm})=J^{\pm},~~~~~\tau(J^3)=J^3,~~~~~\tau(U)=U\nn
&&\tau(j^{\pm})=-j^{\pm},~~~~~\tau (j^{'\pm})=-j^{'\pm},
\eea
which can be used to introduce the following twist in the 
affine superalgebra $A(1,0)^{(1)}$,
\bea
&&(J^{\pm})'_n = J^{\pm}_{n \pm 1},~~~~~(J^3)'_n = J^3_n + 
\frac{k}{2} \delta _{n,0},~~~~~U'_n = U_n\nn
&&(j^{\pm})'_n = j^{\pm}_{n \pm \hf},~~~~~(j^{'\pm})'_n =
j^{'\pm}_{n \pm \hf}.
\label{eq:twist}\eea
Let us extend the superalgebra \eqn{super} by $L_0$, the zero-mode operator
in the Laurent expansion of $T(z)$. It is straightforward to check that the
commutation relations \eqn{super}, together with the extra relations,
\ben
[L_0, \phi_n]=-n \phi_n,~~~\phi_n=J_n^{\pm},J_n^3,U_n,j_n^{\pm},j_n^{'\pm},
\een
are unchanged when one considers the primed operators \eqn{twist} and
\ben
L'_0 = L_0 + J_0^3.
\een
The unprimed superalgebra \eqn{super} with $m,n \in {\bf Z}$ is called the 
Ramond sector of the theory,
while the primed twisted superalgebra is known as the Neveu-Schwarz sector.
The above discussion shows that the two sectors are isomorphic.
The conformal weight, isospin and $U(1)$ charges of physical states are 
related in the following way between the two sectors,
\ben
h^{NS}=h^R + \hf h_-^R,~~~~~~\hf h_-^{NS}=\hf h_-^R+\frac{k}{2},~~~~~
\hf h_+^{NS} = \hf h_+^R.
\een 
{}For definiteness in this paper, 
all our subsequent discussions are in the 
Ramond sector, and we choose the two simple roots to be fermionic (Type I).
The $A(1,0)^{(1)}$ root lattice is generated, in the type I, Ramond picture, 
by three simple roots,
\bea
\alpha _0 &=& (-(\bar{\alpha }_1 + \bar{\alpha }_2),0,1)\nn
\alpha_i &=& (\bar{\alpha }_i, 0,0),~~~~~~~i=1,2
\eea
where $\{\bar{\alpha _1}, \bar{\alpha _2}\}$ are the two fermionic 
roots of $A(1,0)^{(1)}$ introduced in Section 2. The set of positive 
roots, $\Delta ^+$, can be written
as
\ben
\Delta ^+ = (\Delta _0^+ \setminus \tilde{\Delta }_0^+) \cup
 \tilde{\Delta }_0^+ \cup  
(\Delta _1^+ \setminus \tilde{\Delta }_1^+) \cup \tilde{\Delta }_1^+,
\een 
where , in the case of $A(1,0)^{(1)} $, 
$\tilde{\Delta}_0^+ = \Delta _0^+ \setminus \{(0,0,2m)\}$
and $\Delta _1^+ = \tilde{\Delta}_1^+$. One has,
\bea
\tilde{\Delta} _0^+ &=& \{ ( \bar {\alpha}_1+\bar {\alpha}_2),0,m),
( -(\bar {\alpha}_1+\bar {\alpha}_2),0,1+m), (0,0,1+2m), m \in {\bf Z}_+ \},\nn
\tilde{\Delta} _1^+ &=& \{ (\bar{\alpha} _i,0,m ), (-\bar{\alpha} _i,0,1+m ),
m \in {\bf Z}_+ , i=1,2\},
\eea
and therefore,
\ben
\rho = ( 0,h^ {\nu},0).
\een
The dual Coxeter number of $A(1,0)^{(1)}$ is independent of the 
choice of simple roots \cite{kw2} and is $h^{\nu}=1$.

Let us parametrise the highest weight vector by the two quantum 
numbers $h_{\pm}$, corresponding to the eigenvalues of the Cartan 
operators $H_{\pm}=
H_1 \pm H_2$ \eqn{cr}, and by the level $k$ at which the affine 
algebra $A(1,0)^{(1)}$ is considered,
\ben
\Lambda = (\bar {\Lambda }, k,0)=(\hf h_-(\bar{\alpha}_1 + \bar{\alpha}_2)+ 
\hf h_+(\bar{\alpha}_1 - \bar{\alpha_2}), k, 0)
\label{eq:hws}\een
(note that the notion of fundamental weight is ill-defined whenever 
a simple root has zero length).
The different factors in the determinant formula are 
(recall $(\bar{\alpha}_1+\bar{\alpha_2})^2=2$, $m \in {\bf{Z}}_+$) 
and $n \in {\bf Z}_+
\setminus \{0\}$, 
\bea
\tilde{\phi}_n^{(0)}( ((\bar {\alpha}_1+\bar {\alpha}_2),0,m))&=&
h_- + (k+1)m-n\nn
\tilde{\phi}_n^{(0)}(( -(\bar {\alpha}_1+\bar {\alpha}_2),0,1+m))&=&
-h_- + (k+1)(1+m)-n\nn
\tilde{\phi}^{(0)}( (0,0,1+2m))&=& (k+1)(1+2m)\nn
\tilde{\phi}^{(1)}((\bar{\alpha} _i,0,m ))
&=&\hf h_-+(-1)^i\hf h_++(k+1)m, ~~~i=1,2\nn
\tilde{\phi}^{(1)}((-\bar{\alpha} _i,0,1+m ))
&=&-\hf h_--(-1)^i\hf h_++(k+1)(1+m), ~~~i=1,2.\nn
\label{eq:factor}\eea

Our aim is to provide the embedding diagrams and quantum numbers of 
singular vectors within Verma modules built on highest weights 
$\Lambda$ \eqn{hws} whose
quantum numbers   
$h_{\pm},k$ lie
at the intersection of infinitely many lines $\tilde{\phi}_n^{(0)}(\alpha )=
\tilde{\phi}^{(1)}(\alpha ')=0$. This happens when
\ben
k+1= p/q,~~~~~gcd(p,q)=1,
\label{eq:admissible}\een
since one has then,
\bea
\tilde{\phi}_n^{(0)}(( (\bar {\alpha}_1+\bar {\alpha}_2),0,m))&=&
\tilde{\phi}_{n+\nu p}^{(0)}( ((\bar {\alpha}_1+\bar {\alpha}_2),0,m+\nu q))\nn
\tilde{\phi}_n^{(0)}( (-(\bar {\alpha}_1+\bar {\alpha}_2),0,1+m))&=&
\tilde{\phi}_{n+\nu p}^{(0)}(( -(\bar {\alpha}_1+\bar {\alpha}_2),0,1+m+\nu q))
\label{eq:inv}\eea
for $\nu \in  {\bf Z}$. 

Note that for an irreducible highest weight module over $A(1,0)^{(1)}$ to be 
integrable, the conditions on
\ben
m_i = (\Lambda , \alpha_i),~~~i=0,1,2
\een
are \cite{kw2}
\bea
m_1+m_2&=&h_- \in {\bf Z}_+ \setminus \{0\}~~~{\rm or}~~~m_1 =m_2=0, 
{}~{\rm i.e.}~~h_-=h_+=0\nn
m_0&=&k-h_- \in   {\bf Z}_+,
\eea
which corresponds in \eqn{admissible}, to considering $q=1$. 

In order to construct the embedding diagrams, we first encode the 
information
on singular vectors provided by the zeros of the Kac-Kazhdan 
determinant in the following definitions and lemmas. We restrict our analysis to
the case $k+1 \ne 0$.\\

\underline{{\bf Definition 1}} : A singular vector $\chi$ of an $A(1,0)^{(1)}$ 
Verma module is a zero norm vector  
such that   $J_1^-\chi=j_0^+\chi=j_0^{+'}\chi=0$ 
(in the Ramond sector). A highest weight
state is a singular vector whose square length is strictly positive.\\

\underline{{\bf Definition 2}} : A subsingular vector $\Sigma $ of an 
$A(1,0)^{(1)}$ Verma module is a vector such that the
three vectors $J_1^- \Sigma , j_0^+ \Sigma , j_0^{+'} \Sigma $ but
not $\Sigma $ itself can be made to vanish by setting at least one singular
vector to zero in the Verma module. A more mathematically precise
definition can be found in \cite{devos}.
Subsingular vectors are not given by the Kac-Kazhdan formula and are
not included in our diagrams.\\

\underline{{\bf Lemma 1}} : a) If $\chi $ is a singular vector such that 
$L_0 \chi =H \chi$, $J_0^3 \chi =\hf H_- \chi$ and 
$U_0 \chi = \hf H_+ \chi$, and if
$H_- +(k+1)m-n=0$ for some $m \in {\bf Z}_+$ and 
$n \in {\bf Z}_+ \setminus \{0\}$, there exists a singular vector 
corresponding to $\eta =n( (\bar {\alpha}_1+\bar {\alpha}_2),0,m)$ 
with conformal weight
$H+mn$, isospin $\hf H_--n$ and charge $\hf H_+$.

{}~~~~~~~~~~~~~~~~b) If $\chi $ is a singular vector such that 
$L_0 \chi =H \chi$, $J_0^3 \chi =\hf H_- \chi$ and 
$U_0 \chi = \hf H_+ \chi$, and if
$H_- -(k+1)(1+m)+n=0$ for some $m \in {\bf Z}_+$ and 
$n \in {\bf Z}_+ \setminus \{0\}$,
there exists a singular vector corresponding to
$\eta=n( -(\bar {\alpha}_1+\bar {\alpha}_2),0,1+m)$ with conformal 
weight $H+(1+m)n$, isospin $\hf H_-+n$ and charge $\hf H_+$.\\

Lemma 1 allows one to obtain all the uncharged descendants of any singular
vector in an iterative way, by using modified affine Weyl reflexions 
$w_0, w_1$ of the
$\widehat{SU(2)}$ subalgebra of $A(1,0)^{(1)}$. Indeed, if $\chi$ is a 
singular vector
with quantum numbers $H,H_-$ and $H_+$, then\\

(1)the vector $w_0 \chi = [J_{-1}^+]^{k+1-H_-} \chi$ is singular in 
the Ramond sector for $k+1-H_-$ positive integer\\

(2)the vector 
$$
w_1 \chi =[J_0^-]^{H_--1}(2H_- j_0^-j_0^{-'} 
+ [H_+-H_-]J_0^-) \chi 
= 2\{j_0^+, (J_0^-)^{H_-}j_0^-\} \chi = 2\{j_0^{+'}, 
(J_0^-)^{H_-}j_0^{-'} \}\chi $$
{}is singular in the Ramond sector for$H_--1$ positive integer.\\

Whenever the powers $k+1-H_-$ and $H_--1$ are not positive 
integers, the above construction must be 
analytically continued a la Malikov-Feigin-Fuchs. In order to describe this
construction in the specific case of $\hslc$, let us
introduce the vector,
\bea
&&\tilde{w}_0^{(M)} \chi =\nn
&&\prod _{i=1}^{M} \biggl\{~(J_0^-)^{2i(k+1)-H_--1}
                  (~[~-2H_-+4i(k+1)~]~j_0^-j_0^{-'}+~[~H_++H_--2i(k+1)~]~J_0^-)\nn
&&~~~~~~~~~~~~~~~\times
                  (J_{-1}^+)^{(2i-1)(k+1)-H_-}~\biggr\} ~\chi \nn
\label{eq:w0}\eea
with quantum numbers
\bea
H'  &=& H+M^2(k+1)-MH_-\nn
H_-'&=& H_--2M(k+1)\nn
H_+'&=&H_+,
\eea
and the vector 
\bea
&&\tilde{w}_1^{(M)} \chi =\nn
&&\prod _{i=0}^{M-1} \biggl\{~(J_{-1}^+)^{(2i+1)(k+1)+H_-}\nn
&&~~~~~\times (J_0^-)^{2i(k+1)+H_--1}
                  (~[~2H_-+4i(k+1)~]~j_0^-j_0^{-'}+~[~H_+-H_--2i(k+1)~]~J_0^-)~] 
\biggr\}~\chi
\nn
\label{eq:w1}\eea
with quantum numbers
\bea
H'  &=& H+M^2(k+1)+MH_-\nn
H_-'&=& H_-+2M(k+1)\nn
H_+'&=&H_+.
\eea
We also define the vector
\ben
w_0^{(M)} \chi =  (J_{-1}^+)^{(2M+1)(k+1)-H_-}\tilde{w}_0^{(M)} \chi
\een
with quantum numbers
\bea
H'  &=& H+(M+1)^2(k+1)-(M+1)H_-\nn
H_-'&=& -H_-+2(M+1)(k+1)\nn
H_+'&=&H_+,
\label{eq:def3}\eea
and the vector
\bea
\lefteqn{w_1^{(M)} \chi =} \nn
&&(J_0^-)^{2M(k+1)+H_--1}
                 ([2H_-+4M(k+1)]j_0^-j_0^{-'}+[H_+-H_--2M(k+1)]J_0^-)~]
                 \tilde{w}_1^{(M)} \chi \nn
\label{eq:def4}\eea
with quantum numbers
\bea 
H'  &=& H+M^2(k+1)+MH_-\nn
H_-'&=& -H_--2M(k+1)\nn
H_+'&=&H_+.
\eea

In the above expressions, $M$ is a positive integer such that
$M(k+1)\pm H_-$ is a positive integer.
The products are
ordered in such a way that the factor evaluated at $i+1$ 
is at the left of the factor evaluated at $i$.
A remarkable property of $w_0^{(M)}$ and $w_1^{(M)}$ is that their
square is either zero or proportional to the identity, namely,
\ben
(w_0^{(M)})^2 \chi = 
\prod _{i=1}^{M}~ [H_++H_--2(M+1-i)(k+1)]~[H_+-H_-+2(M+1-i)(k+1)] \chi
\label{eq:square1}\een
and 
\ben
(w_1^{(M)})^2 \chi = 
\prod _{i=0}^{M}~ [H_++H_-+2(M-i)(k+1)]~[H_+-H_--2(M-i)(k+1)] \chi.
\label{eq:square2}\een
Similarly, one has,
\ben
\tilde{w}_1^{(M)}~\tilde{w}_0^{(M)}~\chi =
\prod _{i=1}^M (H_++H_--2i(k+1))~(H_+-H_-+2i(k+1))
\label{eq:rel1}\een
and
\ben
\tilde{w}_0^{(M)}~\tilde{w}_1^{(M)}~\chi =
\prod _{i=0}^{M-1} (H_++H_-+2i(k+1))~(H_+-H_--2i(k+1)).
\label{eq:rel2}\een
These properties are easily derived from the definitions \eqn{w0}, 
\eqn{w1}, \eqn{def3},\eqn{def4} and 
the relation,
\ben
(J_0^-)^{-1}(\alpha j_0^{-'}j_0^- + \beta J_0^-)
(J_0^-)^{-1}(\alpha j_0^-j_0^{-'} + \beta J_0^-)= \beta (\beta +\alpha ),
\een
for $\alpha$ and $\beta $ c-numbers.

The analytically continued version of the singular vectors 
$w_0 \chi $ and $ w_1 \chi $ constructed above
is therefore given, when $\chi $ is a singular vector with the quantum
numbers of Lemma 1, by,\\

(1) the singular vector $w_0^{(q-m-1)} \chi $\\

(2) the singular vector $w_1^{(m)} \chi $.\\

The case where $k+1$ is integer corresponds to $q=1, m=0$ and one has
$w_0^{(0)} \chi = w_0 \chi, w_1^{(0)} \chi = w_1 \chi $.\\

\underline{{\bf Lemma 2}} : If $\chi $ is a singular vector such that 
$L_0 \chi =H \chi$, $J_0^3 \chi =\hf H_- \chi$ and 
$U_0 \chi = \hf H_+ \chi$, and if
$H_+-H_- =2(k+1)M $ for some integer $M \geq 0$, there exists a singular vector
corresponding to $\eta = (\bar{\alpha}_1, 0,M)$ with conformal 
weight $H+M$, isospin $\hf H_- -\hf$ and charge $\hf H_+ - \hf$. It is 
given by,
\ben
\xi = \tilde{w}_0^{(M)} j_0^{-'} \tilde{w}_1^{(M)} \chi .
\een

{}Note that for $M=0$, this singular vector is given by $j_0^{-'} \chi$. 
{}For $M=1$, the construction above gives
\bea
\lefteqn{\xi = -4(k+1+H_-)~(k+H_-)\times}\nn
&&~~~~~~\{~(~(k+1)J_0^--j_0^{-'}j_0^-~)j_{-1}^+ - (k+1)H_- j_{-1}^{-'}
  -j_0^{-'} [J_0^-J_{-1}^+-H_-J_{-1}^3 +H_- U_{-1}] ~\} \chi.\nn
\eea
When $k+1+H_-=0$, the singular vector is proportional to 
\bea
&&\{~(~(k+1)J_0^--j_0^{-'}j_0^-~)j_{-1}^+ + (k+1)^2 j_{-1}^{-'}
  -j_0^{-'} [J_0^-J_{-1}^+ + (k+1)J_{-1}^3 -(k+1) U_{-1}] ~\} \chi .\nn
\eea

\underline{{\bf Lemma 3}} :  If $\chi $ is a singular vector such that 
$L_0 \chi =H \chi$, $J_0^3 \chi =\hf H_- \chi$ and 
$U_0 \chi = \hf H_+ \chi$, and if
$H_+-H_- =-2(k+1)(1+M) $ for some integer $M \geq 0$, there exists a 
singular vector
corresponding to $\eta = (-\bar{\alpha}_1, 0,1+M)$ with conformal 
weight $H+1+M$, isospin $\hf H_- +\hf$ and charge $\hf H_+ + \hf$. It is
given by,
\ben
\xi = w_0^{(M)} j_0^- w_0^{(M)} \chi .
\een

{}Note that for $M=0$, this singular vector is 
$[(k-H_-)j_{-1}^{+'} + j_0^- J_{-1}^+] \chi $.
 
\underline{{\bf Lemma 4}} :  If $\chi $ is a singular vector such that 
$L_0 \chi =H \chi$, $J_0^3 \chi =\hf H_- \chi$ and 
$U_0 \chi = \hf H_+ \chi$, and if
$H_++H_- =2(k+1)(1+M) $ for some integer $M \geq 0$, there exists a singular 
vector
corresponding to $\eta = (-\bar{\alpha}_2, 0,1+M)$ with conformal 
weight $H+1+M$, isospin $\hf H_- +\hf$ and charge $\hf H_+ - \hf$. It is given 
by,
\ben
\xi = w_0^{(M)} j_0^{-'} w_0^{(M)} \chi .
\een

{}For $M=0$, it is given by
\ben
[(H_--k)j_{-1}^+ +j_0^{-'}J_{-1}^+] \chi.
\een

\underline{{\bf Lemma 5}} :  If $\chi $ is a singular vector such that 
$L_0 \chi =H \chi$, $J_0^3 \chi =\hf H_- \chi$ and 
$U_0 \chi = \hf H_+ \chi$, and if
$H_++H_- =-2(k+1)M $ for some integer $M \geq 0$, there exists a singular vector
corresponding to $\eta = (\bar{\alpha}_2, 0,M)$ with conformal 
weight $H+M$, isospin $\hf H_- -\hf$ and charge $\hf H_+ + \hf$. It is given by,
\ben
\xi = \tilde{w}_0^{(M)} j_0^{-} \tilde{w}_1^{(M)} \chi .
\een
 
{}For $M=0$, it is given by $j_0^- \chi$.

\section{Embedding diagrams}

The embedding diagrams we construct describe all singular vectors , given with 
their multiplicity, within a given
Verma module with highest weight state $\Lambda $ taken as bosonic for 
definiteness. An arrow originating at a singular vector
$v_1$ and pointing at a singular vector $v_2$ expresses that $v_2$ is a 
descendant of $v_1$, and that there is no singular vector $v_3$ such that
$v_2$ is a descendant of $v_3$, itself descendant of $v_1$.
We split the embedding diagrams in four classes, described below as classes
I to IV, according to whether the Kac-Kazhdan determinant 
$det F_{\eta}(\Lambda)$ for $\Lambda $ highest weight vector has none 
(class I), one (class II), or two zeros (classes III and IV) in the 
fermionic sector
\bea
&&\tilde{\phi}^{(1)}((\bar{\alpha}_i,0,m))=0,~~~~~i=1,2\nn
&&\tilde{\phi}^{(1)}((-\bar{\alpha}_i,0,1+m))=0~~~~~i=1,2,
\eea
see \eqn{factor}.
The standard technique to obtain the embedding
structure of singular vectors is based on an iteration procedure. First, one
considers the zeros of the determinant formula $det F_{\eta}(\Lambda)=0$
for $\Lambda $ the highest weight state, and uses the five lemmas of 
Section 3  to draw the relevant connecting arrows between the singular
vectors corresponding to these zeros. Not all singular vectors are obtained
from  $det F_{\eta}(\Lambda)=0$ however; one must consider the zeros of
 $det F_{\eta}(\Lambda ')=0$ for any singular vector $\Lambda '$ identified
in the previous stage, and use the lemmas again. This iterative procedure
will produce all singular vectors within a given Verma module. It may however
fail to provide the correct embedding structure in three ways. Indeed,

(1) it will not recognise if a singular vector vanishes identically,

(2) it will not provide a complete set of interrelating arrows between
singular vectors,

(3) it will not give the multiplicity of each singular vector.

It is therefore extremely useful to combine it with the knowledge of
analytic expressions for singular vectors in order to provide a complete
embedding diagram. Let us first illustrate how analytic expressions allow
one to identify vanishing singular vectors.
Let $Z_0'$ be a bosonic highest weight vector whose quantum numbers obey the
relation $h_+-h_-=2(k+1)M$, $ M \ge 0.$ Using Lemma 2, one constructs a 
fermionic
singular vector
\ben
Z_0^{'-}= \tilde{w}_0^{(M)} j_0^{'-} \tilde{w}_1^{(M)} Z_0'
\label{eq:ex}\een
whose quantum numbers satisfy the same relation $H_+-H_-=2(k+1)M$. If one
considers $det F_{\eta} (Z_0^{'-})=0$, Lemma 2 produces the singular vector
\ben
Z_0^{'=}= \tilde{w}_0^{(M)} j_0^{'-} \tilde{w}_1^{(M)} Z_0^{'-}.
\een
However, it identically vanishes, as can be seen by replacing $Z_0^{'-}$
by its expression \eqn{ex} and using the result \eqn{rel1} together with 
the fact that $j_0^{'-}$ is a nilpotent fermionic generator.

We now show how analytic expressions allow to obtain relations between singular
vectors which are not in the determinant formula. Take for instance
a highest weight vector $Z_0'$ with $h_+=h_- \in {\bf Z}_+ \setminus \{0\}$. By Lemma 2, there exists
a fermionic singular vector $Z_0^{'-}=j_0^{'-} Z_0'$ with $H_+=H_-=h_--1$.
By Lemma 1, there also exists a singular vector 
\bea
T_0'=w_1 Z_0'&=& (J_0^-)^{h_--1}(2h_- j_0^-j_0^{'-}) Z_0'\nn
&=& 2h_-~(J_0^-)^{h_--1} j_0^- Z_0^{'-}.
\eea
This shows how $T_0'$ is a descendant of $Z_0^{'-}$, a relation missed by the
standard iterative procedure. We would like to stress at this point that
if $Z_0^{'-}$ were the highest weight state of the Verma module, $T_0'$ would
be a $\em{subsingular}$ vector, in the sense of definition 2, and we would not 
have included it in the embedding diagram. But $T_0'$ becomes a 
singular vector when $Z_0^{'-}$ is considered as a fermionic descendant of
the highest weight state $Z_0'$. So the missing arrows are always connected
to the presence of subsingular vectors in the sense just described. Such a
situation occurs in all embedding diagrams where bosonic and fermionic
singular vectors coexist.
We will stress it again in our discussion of class II.

{}Finally, the multiplicity of singular vectors is usually one, except for one 
particular class of highest weight vectors. We will discuss this issue
in plenty details below, in the context of class IV.

In the following, we concentrate on Verma modules built on 
highest weight vectors $\Lambda$ (see \eqn{hws}) whose 
quantum number $h_-$ obeys the constraint
\ben
h_-+(k+1)m-n=0 \label{eq:cond}
\een
where $m,n$ are two integers such that
\ben
0 \le m \le q-1~~~~~{\rm{and}}~~~~~0 \le n \le p-1, \label{eq:range}
\een
and 
\ben
k+1=p/q,~~~~~~p,q \in {\bf{Z}}_+ \setminus \{0\},~~~~gcd(p,q)=1.
\een  
As explained above, the condition of fractional level $k$ together with 
condition \eqn{cond} are a necessary requirement for the Verma  module 
to possess an infinite number of singular vectors. The embedding diagrams have 
different structures according to whether or not extra conditions are imposed
on the highest weight vector quantum numbers $h_{\pm}$. The invariance \eqn{inv}
of equation \eqn{cond} under the shift
\ben
m \rightarrow m + \nu q,~~~~~n \rightarrow n + \nu p
\een
allows one to choose $m$ in the range \eqn{range}. The integer $n$ can be
 parametrized as
\ben
n= (\rho -1) p + \tilde{n},~~~~~ \rho \ge 1,~~~~~0 \le \tilde{n} \le p-1,
\een
but we restrict our analysis to the value $\rho =1$. It is indeed for this 
value of $\rho$ that the characters of the corresponding irreducible
representations may be written in terms of generalised Theta functions.
However, the condition $\rho =1$ is not sufficient to characterise admissible
representations. {}For instance, the Verma modules whose highest weight vectors
quantum number $h_-$ satisfies \eqn{cond} when $n=0$ (edge of the Kac table)
do not lead to admissible representations. 

As already mentioned above, the Verma modules considered 
here fall into four classes.
If $|\Lambda >$ is the Verma module bosonic highest weight state, with quantum
numbers $h, \hf h_-, \hf h_+$ given by
\ben
L_0 |\Lambda >=h|\Lambda >,~~~~~ J_0^3 |\Lambda >=\hf h_-|\Lambda >,~~~~~
 U_0 |\Lambda >=\hf h_+|\Lambda >,
\een
we have,

\underline{Class I} : $|\Lambda > $ has conformal weight $h \ge 0$, arbitrary
real charge $\hf h_+$ and isospin $\hf h_-$ obeying \eqn{cond}
\ben
h_-+(k+1)m-n=0 ,~~~0 \le m \le q-1,~~~0 \le n \le p-1.
\een
All singular vectors are bosonic descendants of the highest weight state, and
therefore have the same fixed arbitrary real charge $\hf h_+$. 
They are organised in four families labeled by a positive integer $a  \ge 0$, 
with quantum numbers
\bea
Z'_a &:& H_a = h + a^2pq+a(qn-pm),\nn
&&(h_-)_a={\bf{ n+2ap}}-m(k+1)\nn
T'_a &:& H_a = h + mn +a^2pq+a(qn+pm),\nn
&&(h_-)_a={\bf{ -n-2ap}}-m(k+1)\nn
Z_{a+1} &:& H_{a+1} =  h + mn +(a+1)^2pq-(a+1)(qn+pm),\nn
&&(h_-)_{a+1}={\bf{ -n+2(a+1)p}}-m(k+1)\nn
T_{a+1} &:& H_{a+1} =  h +(a+1)^2pq-(a+1)(qn-pm),\nn
&&(h_-)_{a+1}={\bf{ n-2(a+1)p}}-m(k+1).
\label{eq:bosonic}\eea

Note that at the edge of the Kac table, when $n=0$, one has the following
identification,
\ben
Z'_{a+1} \equiv Z_{a+1},~~~~~\rm{and}~~~~~ T'_{a+1} \equiv T_{a+1}.
\een
We refer to this case as the collapsed version of the generic case $n \ne 0$.
The corresponding embedding diagrams (Figure 2a and Figure 2b)
are constructed in an iterative way by 
using the lemmas above. One has, in terms of modified Weyl transformations,
\bea
Z_a'&=& (w_0^{(q-m-1)} w_1^{(m)})^a Z_0'\nn
T_a'&=& (w_1^{(m)} w_0^{(q-m-1)})^a w_1^{(m)} Z_0'\nn
Z_{a+1} &=& (w_0^{(q-m-1)} w_1^{(m)})^a w_0^{(q-m-1)} Z_0'\nn
T_{a+1} &=& (w_1^{(m)} w_0^{(q-m-1)})^{a+1} Z_0'.
\eea

\underline{Class II} : $|\Lambda >$ has conformal weight $h \ge 0$, but the 
charge and isospin obey the following constraints,
\ben
h_-+(k+1)m-n=0~~~~~{\rm{and}}~~h_--h_+=-2(k+1)m'
\een
which implies
\ben
h_-+h_+=2(k+1)(m'-m)+2n.
\een
Here, 
\ben
0 \le m \le q-1,~~~1 \le n \le p-1,~~~m' \in \bf{Z}_+
\een
and 
\ben
m'-m=(\sigma -1)q+\tilde{m},~~~~~ \sigma \in {\bf{Z}_+},~~~~~~ 
0 \le \tilde{m} \le q-1.
\een
The case where $m'$ is a negative integer is totally similar and 
leads to embedding diagrams which are mirror images of the ones presented in
{}Figure 3 and Figure 4.
The singular vectors have charge $\hf h_+$ when they are bosonic descendants 
of the highest weight state, and their quantum numbers are those of 
\eqn{bosonic}. 
The fermionic descendants have charge $\hf H_+-\hf$, and their other
quantum numbers are obtained from \eqn{bosonic} by shifting
\ben
n \rightarrow n-1,~~~~~h \rightarrow h+m'.
\een

The value $n=1$ corresponds to a degenerate (collapsed) situation. 
The embedding diagrams for this case and the case $n \ne 1$ are given
in Figure 3 and Figure 4 when $\tilde{m}=0$. If $\tilde{m} \ne 0$, one must 
distinguish between the cases
when $0 \le m+\tilde{m} \le q-1$ and $q \le m+\tilde{m} \le 2q-1$. 
However, the diagrams have the same structure as in the $\tilde{m}=0$ case. 
The only difference is in the singular vector sitting at the annihilation
node in the fermionic sector.

Unlike Class I, Class II possesses bosonic and fermionic singular vectors in 
the same Verma module. The nilpotency of the fermionic generators
in $A(1,0)^{(1)}$ has a crucial impact on the way the singular vectors  are
related. In Figure 3 for instance, the singular vector $T_{\sigma}$ is not
a descendant of $T'_{\sigma-1}$ : the  \lq path\rq between these two vectors is
formally given by
\ben
T_{\sigma} = (w_1^{(m)}w_0^{(q-m-1)})^{\sigma} w_1^{(m)}
             (w_0^{(q-m-1)}w_1^{(m)})^{\sigma-1} T'_{\sigma -1},
\een
but one also has $T'_{\sigma-1}= w_1^{(m)} Z'_{\sigma -1}$. It can be shown that
$w_1^{(m)} w_1^{(m)}Z'_{\sigma -1}$ is zero, using \eqn{square2}.

In order to connect singular vectors of charge $h_+$ to singular vectors of
charge $h_+-1$, one uses the transformations $w_0^{(q-m-1)}$ and $w_1^{(m)}$
as well as 
two fundamental fermionic transformations.
The first one relates $Z'_{\sigma-1}$ and $Z_{\sigma-1}^{-'}$
using Lemma 2, namely,
\ben
Z_{\sigma-1}^{-'}= \tilde{w}_0^{(m+\tilde{m})}~j_0^{-'}~
                 \tilde{w}_1^{(m+\tilde{m})}Z'_{\sigma-1},
\een
which reduces to $Z_{\sigma-1}^{-'}=j_0^{-'}Z'_{\sigma-1}$ when 
$m+\tilde{m}=0$. The second one is not of the kind given in the lemmas. 
Although it connects two singular vectors which correspond to zeros of the
Kac determinant, the latter does not encode the fact that one is the descendant
of the other. This second basic fermionic transformation relates
$Z_{\sigma-1}^{-'}$ and $T'_{\sigma -1}$ as follows,
\ben
T'_{\sigma -1} = \tilde{w}_1^{(\tilde{m})} (J_0^-)^{h_+-1} j_0^-
                \tilde{w}_1^{(m+\tilde{m})} Z_{\sigma -1}^{-'}.
\een 

\underline{Class III} :  $|\Lambda >$ has conformal weight $h \ge 0$, but 
the charge and isospin obey the following constraints,
\ben
h_-+(k+1)m=0~~~~~\rm{and}~~h_--h_+=-2(k+1)m'
\een
which implies
\ben
h_-+h_+=2(k+1)(m'-m).
\een
Here, 
\ben
0 \le m \le q-1,~~~m' \in \bf{Z}_+,
\een
and 
\ben
m'-m=(\sigma -1)q+\tilde{m} \ge 1,~~~~~ \sigma \in {\bf{Z}_+},~~~~~ 
0 \le \tilde{m} \le q-1.
\een
The case where $m'$ is a negative integer produces embedding diagrams 
which are mirror images of the diagrams in Figure 5 and Figure 6. 
The bosonic singular vectors have charge $\hf h_+$, with the other 
quantum numbers given by \eqn{bosonic} when $n=0$,
\bea
Z'_a &:& H_a = h + a^2pq-apm,\nn
&&(h_-)_a={\bf{2ap}}-m(k+1)\nn
T'_{a+1} &:& H_{a +1}= h +(a+1)^2pq+(a+1)pm,\nn
&&(h_-)_{a+1}={\bf{-2(a+1)p}}-m(k+1),
\label{eq:collapsed}\eea
with $a \ge 0$.
The fermionic singular vectors have charge $\hf h_+ -\hf$, with quantum numbers
\bea
Z^{'-}_a &:& H_a = h + m'-m +a^2pq+a(q-pm),\nn
&&(h_-)_a={\bf{ 1+2ap}}-m(k+1)\nn
T^{'-}_a &:& H_a = h + m' +a^2pq+a(q+pm),\nn
&&(h_-)_a={\bf{ -1-2ap}}-m(k+1)\nn
Z^-_{a+1} &:& H_{a+1} =  h + m' +(a+1)^2pq-(a+1)(q+pm),\nn
&&(h_-)_{a+1}={\bf{ -1+2(a+1)p}}-m(k+1)\nn
T^-_{a+1} &:& H_{a+1} =  h +m'-m+(a+1)^2pq-(a+1)(q-pm),\nn
&&(h_-)_{a+1}={\bf{ 1-2(a+1)p}}-m(k+1).
\label{eq:fermionic}\eea
The two diagrams in Figure 5 and Figure 6 correspond to the cases 
$p = 1$ and $p \ne 1$
respectively, with $\tilde{m}=0$. If $\tilde{m} \ne 0$, one must, as in 
Class II, 
distinguish between the cases $0 \le m + \tilde{m} \le q-1$ and 
$q \le m+ \tilde{m} \le 2q-1$. However, the diagrams have the same structure as the ones
given here, and we omit them.

\underline{Class IV} :  $|\Lambda >$ has conformal weight $h \ge 0$, 
but the charge and isospin obey the following constraints,
\ben
h_-+(k+1)m=0~~~~~\rm{and}~~h_--h_+=-2(k+1)m'
\een
which implies
\ben
h_-+h_+=2(k+1)(m'-m).
\een
Here, 
\ben
0 \le m \le q-1,~~~m' \in \bf{Z}_+,
\een
but 
\ben
m'-m=(\sigma -1)q+\tilde{m} \le 0,~~~~~ \sigma \in {\bf{Z}_+}, 
0 \le \tilde{m} \le q-1.
\een
The bosonic singular vectors have charge $\hf h_+$, with the other 
quantum numbers given by \eqn{collapsed}. The fermionic singular vectors 
have either
charge $\hf h_+-\hf$ or $\hf h_+ +\hf$. 

Their other quantum numbers are 
respectively, 
\bea
Z^-_{a+1} &:& H_{a+1} =  h + m' -m+(a+1)^2pq+(a+1)(q-pm),\nn
&&(h_-)_{a+1}={\bf{ 1+2(a+1)p}}-m(k+1)\nn
T^-_a &:& H_a =  h +m'+a^2pq+a(q+pm),\nn
&&(h_-)_a={\bf{ -1-2ap}}-m(k+1).
\eea
and
\bea
Z^+_{a+1} &:& H_{a+1} =  h -m'+(a+1)^2pq+(a+1)(q-pm),\nn
&&(h_-)_{a+1}={\bf{ 1+2(a+1)p}}-m(k+1)\nn
T^+_a &:& H_a =  h +m-m'+a^2pq+a(q+pm),\nn
&&
(h_-)_a={\bf{ -1-2ap}}-m(k+1)
\eea
with $a \ge 0$. The corresponding embedding diagram is given in Figure 7.

The double multiplicity of the vectors $T_{a+1}'$ and $Z_{a+2}'$ for $a \ge 0$
is a rather new and remarkable feature. Until recently (\cite{Doer},
\cite{devos}), it was common 
belief that the singular vectors appearing in embedding diagrams all had 
multiplicity one. Our analysis for $\hslc$ confirms the presence of 
singular vectors of higher multiplicities for particular choices of highest
weight state quantum numbers (namely class IV). We can indeed 
generalise the MFF construction in a highly nontrivial way in order to
properly take into account the complications due to the presence of nilpotent
fermionic generators.  
Given the importance of the higher multiplicities of singular vectors,
in particular in deriving character formulas, 
we now illustrate our technique and construct two singular vectors $T_1^{'(1)}$
and $T_1^{'(2)}$ with the quantum numbers of $T_1'$ as they appear in 
\eqn{collapsed}.

We restrict ourselves to the case $m' < m/2 $, $m$ odd. Similar ideas can be
used for $m$ even, and/or any $m'$ such that $m'-m \le 0$. The vector 
$T_1^{'(1)}$ is easily constructed as 
\ben
T_1^{'(1)}= w_1^{(m)}~w_0^{(q-m-1)} Z_0',
\een 
with the help of the generalised Weyl transformations introduced in the
previous section. 
The second vector is far from being trivial as a descendant of the highest
state $Z_0'$. A reasonable starting point would be to construct
the state $w_1^{(m)}~Z_0'$. However, with the class IV choice of
quantum numbers for $Z_0'$ in particular, this state identically vanishes
due to its internal fermionic structure, as can be checked by using the 
definition \eqn{def4}. One therefore needs to \lq improve\rq  the state
$w_1^{(m)}~Z_0'$ to avoid its vanishing. As described in the following 
expressions, the improved state, called $~\widehat{w_1}^{(m)} Z_0'$,
is given by a linear combination of the
appropriately dressed neutral objects
${\rm {log}} J_{-1}^+$  and $(J_0^-)^{(-1)}j_0^-j_0^{'-}$. Explicitly,
the singular vector $T_1^{'(2)}$ is given by
\ben
T_1^{'(2)}= w_1^{(m)}~w_0^{(q-m-1)}~\widehat{w_1}^{(m)} Z_0',
\een   
with
\bea
\lefteqn{\widehat{w_1}^{(m)}~Z_0'=}\nn
&&4h_+^2~\tilde{w}_0^{(m')}
\biggl {[}(J_0^-)^{-h_+-1}j_0^{'-}j_0^-~~\tilde{w}_1^{(m/2-m'-1/2)}~~
\alpha  {\rm{log}} J_{-1}^+
\tilde{w}_0^{(m/2-m'-1/2)}~(J_0^-)^{h_+-1}j_0^{-}j_0^{'-}~\nn
&&~~~~~~~~~~~~~~~+\beta (J_0^-)^{-1}j_0^-j_0^{'-}~\biggr {]}~
\tilde{w}_1^{(m')}~Z_0'
\label{eq:what}\eea
where
$\alpha = h_+$ and 
\ben
\beta = \prod _{j=1}^{m/2-m'-1/2}(h_++(2j-1)(k+1))~ (h_+-(2j-1)(k+1)).
\een
The function
${\rm {log}}J_{-1}^+$ allows to write up vectors such as $T_1^{'(2)}$ in a 
concise
way, making it easy to check they are singular. Indeed, 
$J_1^-$ commutes with $\tilde{w}_0^{(M)}$ and $\tilde{w}_1^{(M)}$ for
any value of $M$, and because the commutator
\ben
[ J_1^-, {\rm log}J_{-1}^+] = (J_{-1}^+)^{-1}(k+1-2J_0^3)
\een
vanishes when evaluated on a state whose
quantum number $H_-$ satisfies $k+1-H_-=0$ (it is always the case
in our construction), one concludes that, in 
particular, $J_1^- T_1^{'(2)}=0$. Although
$j_0^+$ and $j_0^{'+}$ commute with $\log J_{-1}^+$, it is interesting to 
note that checking $j_0^+ T_1^{'(2)}=j_0^{'+}T_1^{'(2)} =0$ is not 
straightforward. Indeed, $\log J_{-1}^+$ is 
not an eigenvector of $J_0^3$ (although \eqn{what} is), i.e.,
\ben
[J_0^3, {\rm log}J_{-1}^+]~=~1,
\een
and hence the corrective term 
$\beta (J_0^-)^{-1}j_0^-j_0^{'-}$ is introduced in \eqn{what} in order to 
ensure that $j_0^+ T_1^{'(2)}=0$.

Although $\log J_{-1}^+$ appears in the formal expression of some of our 
singular vectors, it does actually not survive in any evaluation of the
vector, since it can be commuted through with the help of the following
relations,
\bea
\phantom{~} [ j_0^-, {\rm log}J_{-1}^+ ] & = & j_{-1}^{'+} (J_{-1}^+)^{-1},\nn
\phantom{~}   [ j_0^{'-}, {\rm log}J_{-1}^+ ] & = & -j_{-1}^{+} (J_{-1}^+)^{-1},\nn
\phantom{~} [ J_0^-, {\rm log}J_{-1}^+] & = & -2 (J_{-1}^+)^{-1} J_{-1}^3
+ (J_{-1}^+)^{-2}J_{-2}^+,
\eea
and disappears because the first term in the square bracket of \eqn{what}
vanishes identically when $\log J_{-1}^+$ is removed.

We have concentrated here on the case where $m$ is odd. If $m$ is even, 
one substitutes appropriately the expression 
$[h_+\log J_0^- + (J_0^-)^{-1}j_0^-j_0^{'-}]$ for $\log J_{-1}^+$.

We now proceed to indicate the relation between the fermionic 
singular vector $T_0^+$ and the two uncharged singular vectors $T_1^{'(1)}$
and $T_1^{'(2)}$. 
The construction given in Lemma 5 for $T_0^+$, namely,
\ben
T_0^+= \tilde{w}_0^{(m-m')}~j_0^-~\tilde{w}_1^{(m-m')}Z_0',
\een
leads to a state which vanishes identically in this class, again because of 
its internal fermionic nature. 
The actual fermionic vector is obtained by using an \lq improved\rq version
of Lemma 5, inspired by the idea above, namely,
\bea
&&T_0^+=
2h_+\tilde{w}_0^{(m-m')}
\biggl {[}j_0^-~~\tilde{w}_1^{(m/2-m'-1/2)}~~
\alpha  {\rm {log}} J_{-1}^+
\tilde{w}_0^{(m/2-m'-1/2)}~
(J_0^-)^{h_+-1}j_0^{-}j_0^{'-}~\nn
&&~~~~~~~~~~~~~~~~~~~~~~~~~~~-\beta (J_0^-)^{h_+}j_0^-~\biggr {]}~
\tilde{w}_1^{(m')}~Z_0',
\nn \eea
with $\alpha$ and $\beta $ given as before.
It is now almost straightforward to identify which particular linear 
combination
of $T_1^{'(1)}$ and $T_1^{'(2)}$ is a descendant of
$T_0^+$,
\bea
\lefteqn{N_1(-4h_+^2\beta N_2T_1^{'(1)}+T_1^{'(2)})=}\nn
&&w_1^{(m)}~w_0^{(q-m-1)}~\tilde{w}_0^{(m')}
(J_0^-)^{-h_+-1}(-2h_+j_0^{'-})~\tilde{w}_1^{(m-m')}~T_0^+,\nn
\eea
where $N_1$ is a nonzero normalisation constant given by \eqn{rel1} for 
$M=m'-m$, $H_+=-H_-=h_++1$, and $N_2$ is similarly given by \eqn{rel2} for
$M=m'$, $H_-=h_-$ and $H_+=h_+$.

{}Finally, the uncharged singular vector $T_1^{'(2)}$ can be seen as a descendant
of the fermionic singular vector $T_0^-$ in the following way. By lemma 2,
one constructs $T_0^-$ as,
\ben
T_0^- = \tilde{w}_0^{(m')}j_0^{'-}\tilde{w}_1^{(m')}Z_0'
\een
and
\bea
\lefteqn{N'T_1^{'(2)}=}\nn
&&4h_+^2~\tilde{w}_0^{(m')}
\biggl {[}(J_0^-)^{-h_+-1}j_0^{'-}j_0^-~~\tilde{w}_1^{(m/2-m'-1/2)}~~
\alpha  {\rm{log}} J_{-1}^+
\tilde{w}_0^{(m/2-m'-1/2)}~(J_0^-)^{h_+-1}j_0^-~\nn
&&~~~~~~~~~~~~~~~+\beta (J_0^-)^{-1}j_0^-~\biggr {]}~
\tilde{w}_1^{(m')}~T_0^-.
\eea
The normalisation factor $N'$ is,  
\ben
N'= \prod_{i=0}^{m'-1}(h_++h_-+2i(k+1)-2)~(h_+-h_--2i(k+1)).
\een

Once more, this detailed analysis illustrates very well the power of our 
analytic expressions in 
relating singular vectors between themselves within a given
embedding diagram.

\section{Conclusions}
The Lie superalgebra $A(1,0)$ and its affinisation $A(1,0)^{(1)}$
play a crucial role in the
description of noncritical $N=2$ superstrings. In order to study the space
of physical states of the latter theory, using the tool provided by topological
$G/G$ WZNW models, a detailed analysis of various modules over $A(1,0)^{(1)}$
is needed.
Many Lie superalgebras share with $A(1,0)$ the property that two sets of 
simple roots may not be equivalent up to Weyl tranformations, which are 
generated by reflections with respect to bosonic simple roots. An added 
technical complication in $A(1,0)$ is the fact that the fermionic roots are 
lightlike,
which prevents one from defining coroots and fundamental weights in a
straightforward way. These properties are emphasized in Section 2.
The classical and quantum free field Wakimoto representations of $\hslr$ 
built with two inequivalent sets of simple roots are 
given in \cite{us2}. It is shown there that there exists
a set of field transformations which relate the two Wakimoto representations
in the classical and the quantum case.
Section 3 organises the information provided by the Kac-Kazhdan determinant
formula relevant to the Lie superalgebra $A(1,0)^{(1)}$
in five lemmas. The Malikov-Feigin-Fuchs construction is generalised to
incorporate transformations which relate bosonic and fermionic singular
vectors within a Verma module.
Section 4 provides vital information for the construction of admissible
representations of $A(1,0)^{(1)}$, namely the quantum numbers and embedding
diagrams of the singular vectors appearing in highest weight Verma modules
when the level $k$ of the algebra satisfies the necessary condition
$k+1 =p/q$ with $p,q$ nonzero positive relatively prime integers. We believe 
that the extra conditions leading to the seven embedding diagrams of Classes
I,II,III, IV are sufficient to determine all admissible representations, whose
characters should provide  finite representations of the modular group.
Our analysis clearly shows a very close link between the embedding diagrams
of Section 4 and those of some completely degenerate representations of the
$N=2$ superconformal algebra \cite{Doer}. This striking similarity is 
reminiscent of the link between the admissible $\widehat{sl(2;{\bf C})}$ 
modules and
the degenerate Virasoro modules, and between the admissible
$\widehat{osp(1/2;{\bf C})}$ modules
and the $N=1$ degenerate $N=1$ superconformal modules. A recent paper 
\cite{semikhatov} offers some explanation of this similarity.

{\bf Acknowledgements}

We would like to thank Gerard Watts for sharing with us his insight on 
$N=2$ singular vectors, and for making one of his symbolic computer programmes 
available to us. We also thank Alexei Semikhatov for discussions on the
relation between $N=2$ and $A(1,0)^{(1)}$ singular vectors, and
Jonathan Evans for pointing out reference \cite{k1}.
Anne Taormina acknowledges the U.K. Engineering and Physical Sciences
Research Council for the award of an Advanced Fellowship. She also thanks
Cern for its hospitality, where part of this work was done.

\begin{figure}[h]
\begin{center}
\leavevmode
\epsfysize=17truecm
\epsffile{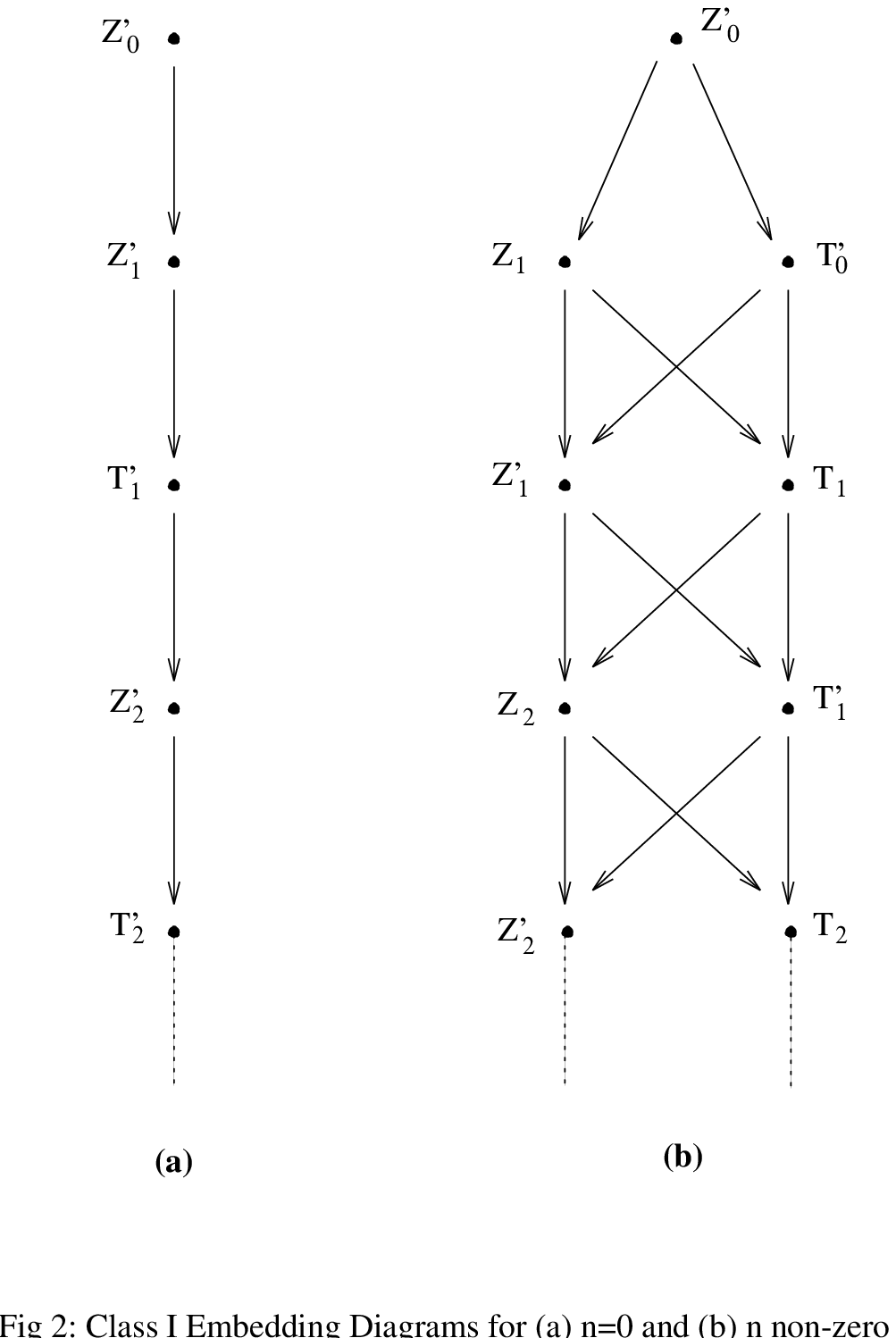}
\end{center}
\end{figure}
\vfill
\eject

\begin{figure}[h]
\begin{center}
\leavevmode
\epsfysize=22truecm
\epsffile{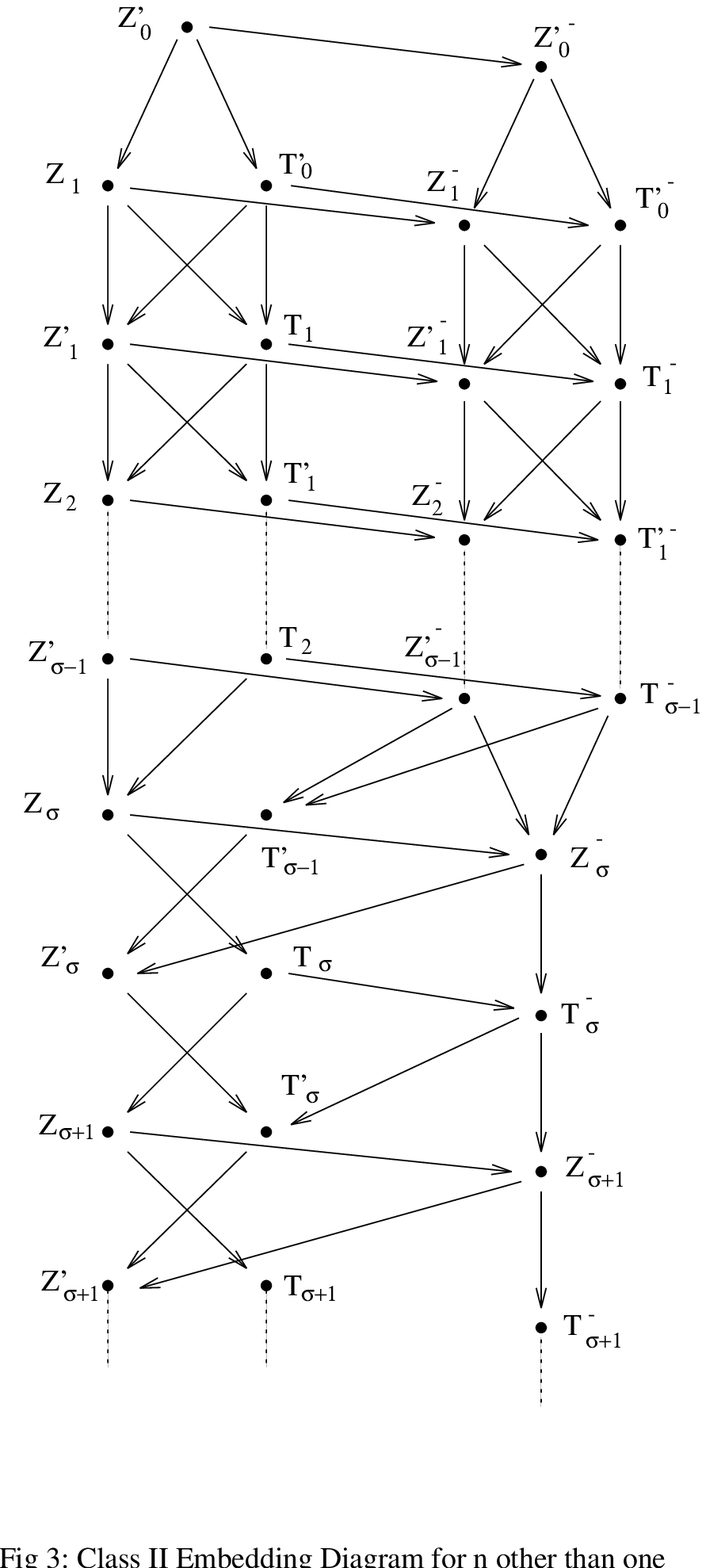}
\end{center}
\end{figure}
\vfill
\eject

\begin{figure}[h]
\begin{center}
\leavevmode
\epsfysize=22truecm
\epsffile{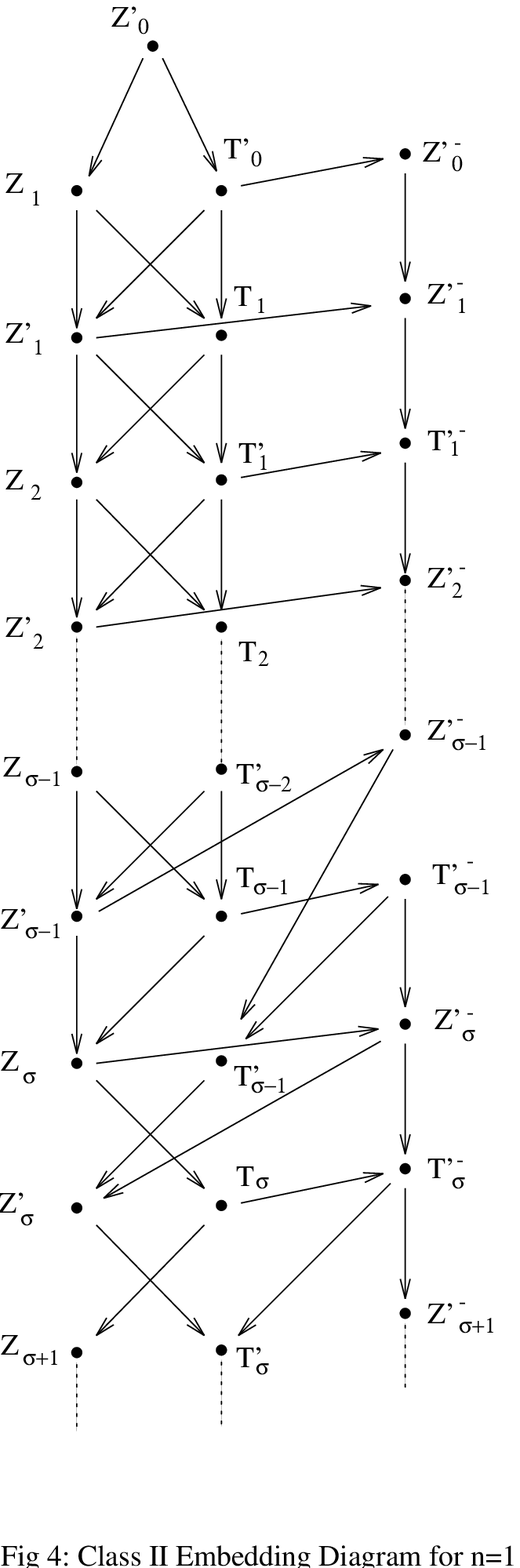}
\end{center}
\end{figure}
\vfill
\eject

\begin{figure}[h]
\begin{center}
\leavevmode
\epsfysize=20truecm
\epsffile{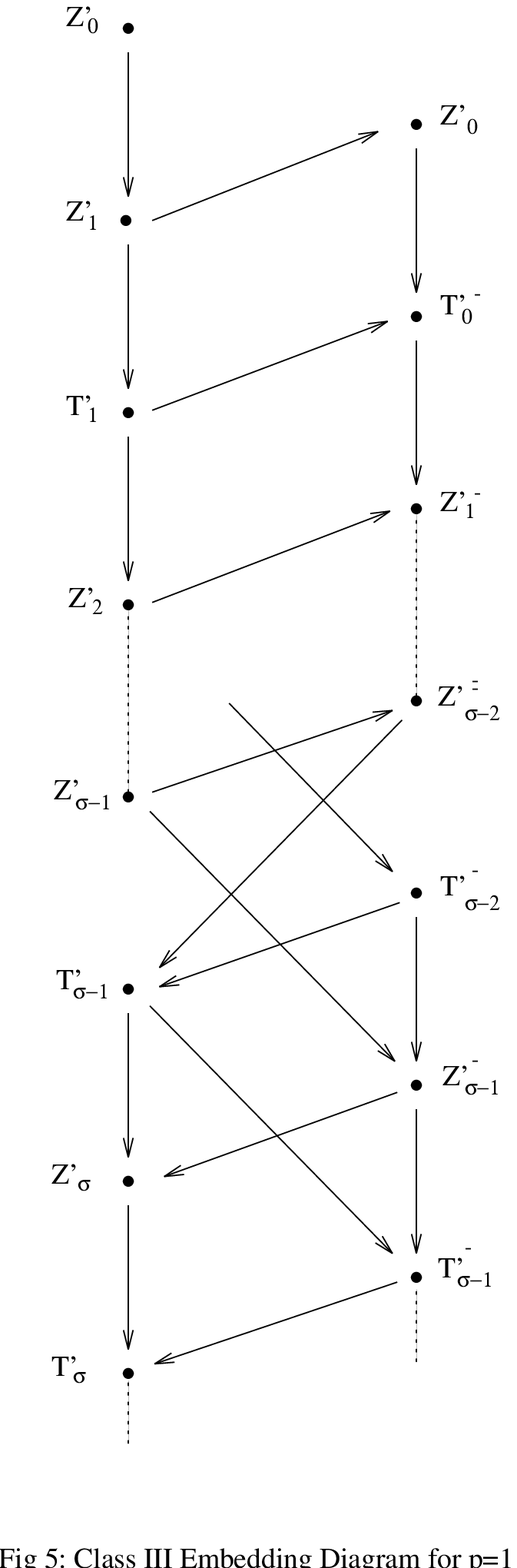}
\end{center}
\end{figure}
\vfill
\eject

\begin{figure}[h]
\begin{center}
\leavevmode
\epsfysize=22truecm
\epsffile{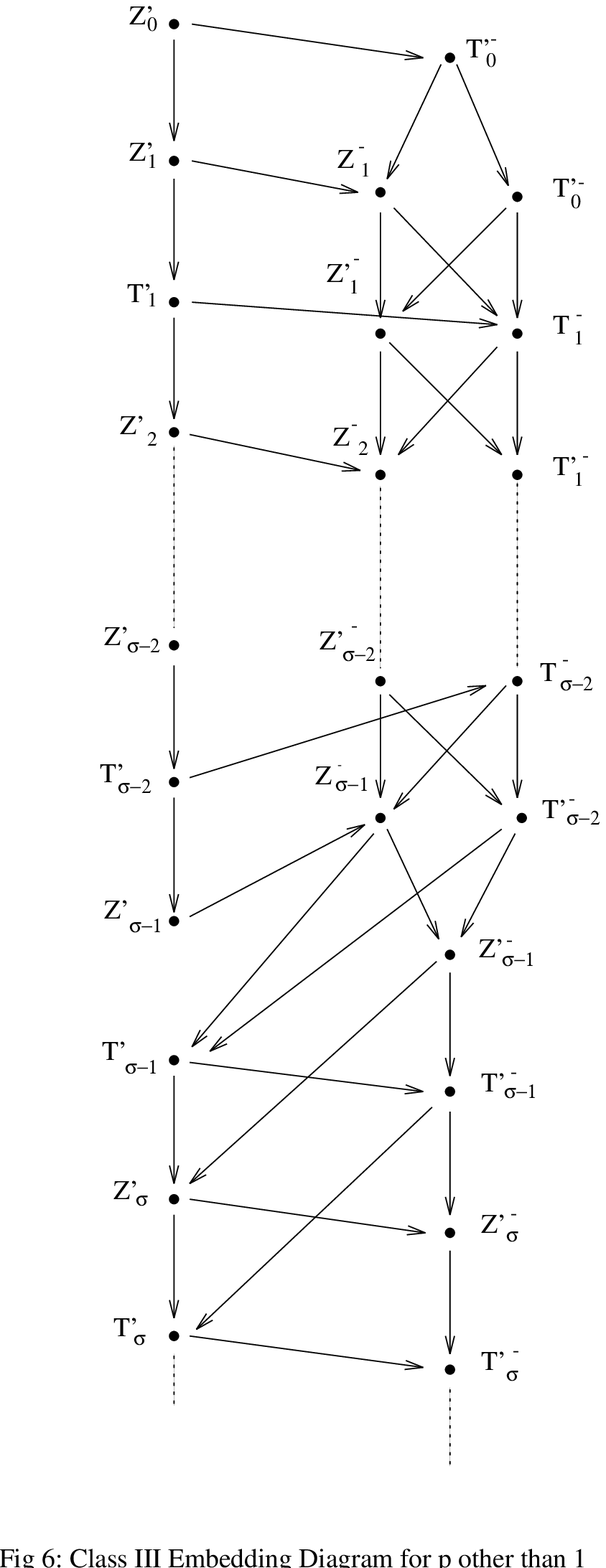}
\end{center}
\end{figure}
\vfill
\eject

\begin{figure}[h]
\begin{center}
\leavevmode
\epsfysize=17truecm
\epsffile{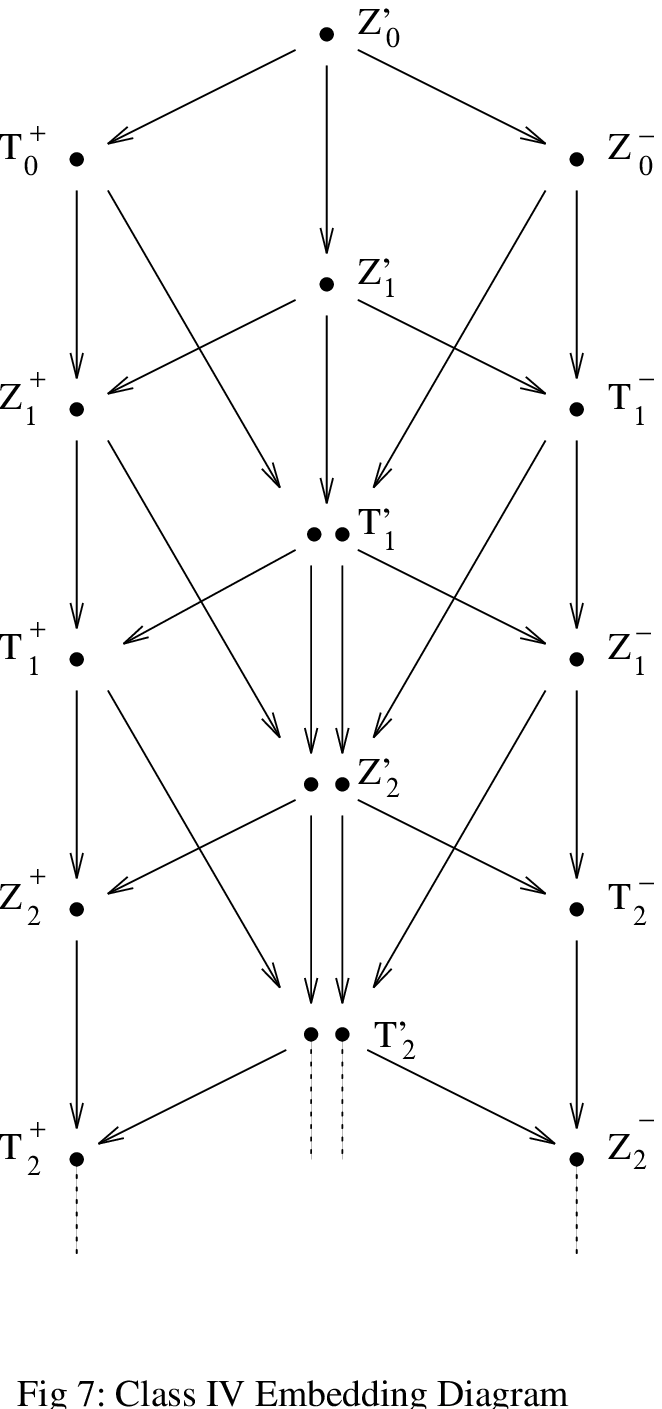}
\end{center}
\end{figure}
\vfill
\eject

\end{document}